\documentclass[useAMS,usenatbib,usegraphicx]{mn2e} 
 
\usepackage{mdwtab}
\usepackage{amssymb}
\usepackage{amsfonts}
\usepackage{amsmath}
\newcommand{\new}[1]{\textrm{#1}}
 
\title[The Tidal Tails of 47 Tucanae]{The Tidal Tails of 47 Tucanae} 
\author[Richard R. Lane]{Richard R. Lane$^{1}$\thanks{E-mail: 
rlane@astro-udec.cl (RRL); akuepper@astro.uni-bonn.de (AHWK);
dcheggie@ed.ac.uk (DCH)}, Andreas H.W. K\"upper$^{2,3}$ and Douglas C. Heggie$^{4}$\\ 
$^{1}$Departamento de Astronom\'ia, Universidad de Concepci\'on,
  Casilla 160 C, Concepci\'on, Chile\\ 
$^{2}$Argelander-Institut f\"ur Astronomie (AIfA), Auf dem H\"ugel
  71, 53121 Bonn, Germany\\
$^{3}$European Southern Observatory, Alonso de Cordova 3107, Vitacura, Santiago, Chile\\
$^{4}$University of Edinburgh, School of Mathematics and Maxwell
  Institute for Mathematical Sciences, King's Buildings, Edinburgh EH9
  3JZ, U.K.} 
\begin{document} 
 
\date{Accepted..... Received.....; in original form.....} 
 
\pagerange{\pageref{firstpage}--\pageref{lastpage}} \pubyear{2002} 
 
\maketitle 
 
\label{firstpage} 
 
\begin{abstract} 
The Galactic globular cluster 47 Tucanae (47 Tuc) shows a rare increase in its
velocity dispersion profile at large radii, indicative of energetic, yet
bound, stars at large radii dominating the velocity dispersion and,
potentially, of ongoing evaporation. Escaping stars will form tidal tails, as
seen with several Galactic globular clusters, however, the tidal tails of 47
Tuc are yet to be uncovered. We model these tails of 47 Tuc using the most
accurate input data available, with the specific aim of determining their
locations, as well as the densities of the epicyclic overdensities within the
tails. The overdensities from our models show an increase of 3-4\% above the
Galactic background and, therefore, should be easily detectable using matched
filtering techniques. We find that the most influential parameter with regard
to both the locations and densities of the epicyclic overdensities is the
Heliocentric distance to the cluster. Hence, uncovering these tidal features
observationally will contribute greatly to the ongoing problem of determining
the distance to 47 Tuc, tightly constraining the distance of the cluster
independent of other methods. Using our streakline method for determining the
locations of the tidal tails and their overdensities, we show how, in
principle, the shape and extent of the tidal tails of any Galactic globular
cluster can be determined without resorting to computationally expensive
$N$-body simulations.
\end{abstract} 
 
\begin{keywords} 
Galaxy: globular clusters: general -- Galaxy: globular clusters: individual:
47 Tucanae -- Galaxy: kinematics and dynamics -- methods: numerical
\end{keywords} 
 
\section{Introduction}\label{intro}

47 Tucanae (47 Tuc) is among the most interesting and best studied globular
clusters of the Milky Way (MW). It is one of the most massive, and exhibits
some peculiar properties, including a recently detected rise in the velocity
dispersion at large cluster radii \cite[][]{Lane10a}. Such a rise in the
velocity dispersion profile of a globular cluster was first noted some time
ago by \cite{Drukier98} in M15. While a rise of this nature is intuitively
unexpected for a velocity dispersion profile in Newtonian dynamics, it arises
naturally in modified gravitational theories like MOND
\cite[e.g.][]{Scarpa10}.

In the case of 47 Tuc \cite{Lane10b} posed, as an alternative explanation for
this phenomenon, the existence of two separate kinematic populations, a relic
of its formation. A third explanation was suggested more recently by
\cite{Kupper10b}, who found a similar rise in velocity dispersion in their
$N$-body models of star clusters orbiting in galactic tidal fields. Their
alternate explanation arises from the contribution of energetically unbound
stars which haven't yet escaped the cluster potential. These ``potential
escapers'' of \cite{Kupper10b} are preferentially located at larger radii,
where they dominate the velocity dispersion. Moreover, they are an indicator
of ongoing, relaxation driven mass loss (evaporation), since a star has to
become a potential escaper before it eventually escapes from the cluster to
become a member of the tidal tails
\cite[][]{Fukushige00,Kupper10b}. Dynamically cold tidal tails emanating from
47 Tuc are, therefore, to be expected in this scenario. Hence, locating these
tails will help to clarify the mechanism responsible for rising velocity
dispersion profiles in globular clusters.

Tidal streams formed in this way have been observed for other MW globular
clusters like Palomar 5 and NGC 5466 \cite[][]{Odenkirchen03,Grillmair06}, but
no tidal tails associated with 47 Tuc have yet been uncovered. However,
\cite{Lane10a} found 25 stars outside the tidal radius given in the
\cite{Harris96} catalogue, possible evidence for evaporation. Furthermore,
these apparently extra-tidal stars precisely follow the trend of increased
velocity dispersion as found by \cite{Kupper10b} in $N$-body computations
\cite[see Fig.~11 by][]{Lane10a}.

The lack of prominent extra-tidal features could be due to the non-linear
motion of stars within tidal tails. \cite{Kupper08} and \cite{Just09} found
that the epicyclic motion of evaporating stars results in periodic over- and
underdensities within the tidal tails, even if they are formed by a constant
stream of escaping stars. The periodicity is spatial, meaning that the
overdensities are regularly spaced, at least in the case of a circular
galactic orbit, and are found in both the leading and trailing tails. Their
spatial separation equates to the distance through which an escaper drifts
(relative to the cluster) in one epicyclic period, and decreases slowly as the
cluster loses mass. Close to the tidal radius the escaping stars are
accelerated away from the cluster, therefore, this motion leads to a
pronounced underdensity of stars close to the cluster. Hence, most clusters
should not exhibit extended tidal features, but rather a steeply decreasing
surface brightness close to the cluster (e.g. see
Fig.~\ref{densityprofile}). The epicyclic overdensities, on the other hand,
can reach surface densities several times higher than the average density
within the tidal tails, but are typically located many tidal radii from the
cluster and thus the association with the cluster may not be obvious. The
positions of these overdensities can be accurately predicted, however, if the
dynamical state of the cluster (i.e.~its orbit, location \& mass) is
sufficiently constrained \cite[][]{Kupper10a, Kupper12}.

Since the orbit of 47 Tuc has been determined in a number of observational
studies, and it has been subject to extensive numerical modelling, its
dynamical state is thought to be well understood. Thus, our knowledge of 47
Tuc provides an opportunity for predicting and finding the overdensities in
its tidal tails and, therefore, constraining the dynamical state of 47 Tuc
independently from other methods. Moreover, the detection of the overdensities
in the tails of 47 Tuc will shed light on the consistency of our theoretical
understanding of Newtonian gravity.

This work is organised as follows: based on the most accurate data available
(Sec.~\ref{data}), we calculate the orbital parameters of 47 Tuc and predict
the shape of its tidal tails and, most importantly, the positions and extents
of the epicyclic overdensities (Sec.~\ref{results}). Finally we give a short
summary and conclusions in Sec.~\ref{conclusions}.
 
\section{Input data}\label{data}

For this work we require accurate input data to ensure the best quality
orbital model (see Table~\ref{modelparams} for a summary). For this reason we
have chosen to take our positional information from \cite{McLaughlin06} who
claim a nearly complete, uniform sample of all stars within $1.\!\!'5$
($\sim5$ core radii) of the cluster centre, which the authors calculate to be
at
(RA,dec)=($00^{\rm{h}}24^{\rm{m}}05.\!\!^{\rm{s}}67\pm0.\!\!^{\rm{s}}07$,$-72^\circ04^{\rm{'}}52.\!\!''62\pm0.\!\!''26$).
We also take our Heliocentric distance information from \cite{McLaughlin06},
who calculate a kinematic distance of $4.02\pm0.35$\,kpc by comparing the
velocity dispersion of stars within the inner $\sim105''$ with the proper
motions in the same region to estimate a distance to the cluster.

For our proper motion we use the values derived by \cite{Anderson04}, who used
{\it Hubble Space Telescope} ({\it HST}) data to find the proper motion of 47
Tuc to be $5.64\pm0.20$\,mas\,yr$^{-1}$ in RA and
$-2.05\pm0.20$\,mas\,yr$^{-1}$ in dec. These data are considered the best
available data for the current work because the Small Magellanic Cloud is in
the same line of sight as 47 Tuc, allowing the authors to obtain very accurate
differential proper motions. Indeed they claim an order of magnitude greater
accuracy than previous studies.

\begin{figure}
  \begin{centering}
  \includegraphics[width=0.48\textwidth]{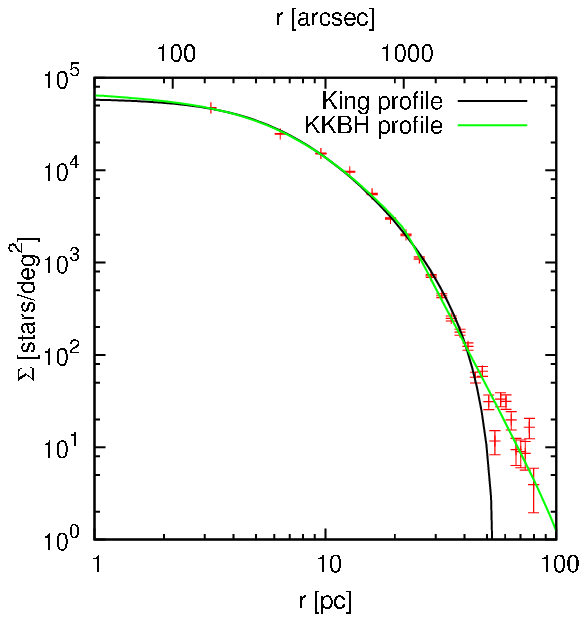}
  \caption{Surface density profile of 47 Tuc based on data from
    \citet{Lane10a}, overlaid with a King profile \citep[][ solid black
      curve]{King62} and a KKBH profile \citep[][ solid green
      curve]{Kupper10b}. The radius where the surface density drops to zero is
    56\,pc in the case of the King profile. The cut-off radius of the core of
    the KKBH profile is slightly larger (65\,pc). However, the power-law
    component of the KKBH extends out to the last data point with a slope of
    $-4.8$.}
  \label{densityprofile}
  \end{centering}
\end{figure}

We have chosen to use the radial velocity of $-16.85\pm0.16$\,km\,s$^{-1}$
derived by \cite{Lane10a} from the largest known spectroscopic GC study to
date. Their survey of $\sim~2200$ individual spectra of stellar members
provides the most accurately determined radial velocity to date. The surface
density profile from the \cite{Lane10a} dataset can be seen in
Fig. \ref{densityprofile}. Note that the King model fit is excellent out to
large radius (where the influence from the potential escapers becomes
important, and apparent) and the King tidal radius from the fit is in
excellent agreement to that given by the \cite{Harris96} catalogue. Fitting a
KKBH profile \citep{Kupper10b} to the same data yields a slightly larger
cut-off radius for the core profile (65\,pc) and a power-law slope of $-4.8$
at large cluster radii. As we show later, the stars at radii larger than the
cut-off radius are not extra-tidal. The theoretical tidal radius of 47 Tuc is
larger than the range of the data (see Sec.~\ref{results}), so the fitted
cut-off radii are both well inside the tidal sphere. Hence, the surface
density profile does not show a clear cut-off at about 60\,pc, but continues
with a constant power-law slope. Extra-tidal signatures like tidal distortions
are only expected at much larger radii of a few hundred parsecs
(e.g. see~Fig.~\ref{warm}).

To estimate surface densities in the modelled tidal tails and their epicyclic
overdensities, it is important to have an independently derived mass-loss rate
for the cluster. Using detailed Monte Carlo models of the dynamical evolution
of 47 Tuc, \cite{Giersz11} found an escape rate of 13.9 stars per Myr, with an
average stellar mass of $0.404\rm\,{M}_\odot$, giving a mass loss rate of
$5.61\rm\,{M}_\odot$\,Myr$^{-1}$. We have adopted their value here.

Many studies exist with measurements of the total mass of 47 Tuc, however, the
most recent of these, \cite{Lane10a} and \cite{Giersz11}, based on
observations and numerical modelling, respectively, find masses of
$1.1\times10^6$\,M$_\odot$ and $0.9\times10^6$\,M$_\odot$ which are in
remarkable agreement. We have, therefore, chosen $1.0\times10^6$\,M$_\odot$ as
the baseline mass for our models.

The Solar galactocentric distance we take from \cite{Gillessen09} who find
$R_0=8.33\pm0.35$\,kpc by fitting orbits to stars associated with the Galactic
centre. Since the distance to the the Galactic centre strongly correlates with
the mass of the central black hole \cite[see Fig.~15 by][]{Gillessen09} these
orbital fits tightly constrain the distance to the Galactic centre,
independent of other methodologies.

In a review of twenty independent measurements of the Solar circular velocity
dating back to 1974, \cite{Kerr86} concluded that a case can be made for
accepting the IAU standard value of 220\,km\,s$^{-1}$, since the mean value of
these twenty studies is 222.2\,km\,s$^{-1}$. We have, therefore, chosen to use
the IAU standard of 220\,km\,s$^{-1}$ for the current study \cite[see
  also][who find $V_{\rm c}=224\pm13$\,km\,s$^{-1}$, in agreement with the IAU
  standard]{Koposov10}.

For the Local Standard of Rest we are adopting ($U,V,W$)$_{\rm
  LSR}=(11.1^{+0.69}_{-0.75},12.24^{+0.47}_{-0.47},7.25^{+0.37}_{-0.36})$\,km\,s$^{-1}$
by \cite{Schonrich10} as this is the only study to take into account the
metallicity gradient in the Galactic disc. Without adjusting for this
gradient, the $V$ component is significantly affected due to the correlation
between stellar group colours and the radial gradients of their properties.

The coordinates of the Galactic North Pole, which are required for coordinate
and proper motion transformations, we take from \cite{Reid04}. All input data
are summarised in Table \ref{modelparams}.

\begin{table*}
\begin{center}
\caption{Input parameters}\label{modelparams}
\begin{tabular}{@{}ccc@{}}
\hline
\hline
Parameter & Value & Reference\\
\hline
$D_{Helio_{\rm 47\,Tuc}}$ & $4.02$\,kpc & \cite{McLaughlin06}\\
$\alpha_{_{\rm 47\,Tuc}}$ (J2000) & $00^{\rm h}24^{\rm m}05.\!\!^{\rm s}67$ & \cite{McLaughlin06}\\
$\delta_{_{\rm 47\,Tuc}}$ (J2000) & $-72^\circ04'52.\!\!''62{\rm }$ & \cite{McLaughlin06}\\
$\mu_\alpha$ & 5.64\,mas\,yr$^{-1}$ & \cite{Anderson04}\\
$\mu_\delta$ & -2.05\,mas\,yr$^{-1}$ & \cite{Anderson04}\\
$V_{\rm r}$ & $-16.85$\,km\,s$^{-1}$ & \cite{Lane10a}\\
\hline
$r_{\rm t}$ & 56\,pc & \cite{Harris96}; this work\\
$M_{\rm tot}$ & $1.0\times10^6$M$_\odot$ & \cite{Lane10a,Giersz11}\\
$\dot{N}$ & 13.9\,Myr$^{-1}$ & \cite{Giersz11}\\
$\dot{M}$ & 5.61\,M$_\odot$\,Myr$^{-1}$ & \cite{Giersz11}\\
$\dot{M}/\dot{N}$ & $0.404\,M_\odot$ & \cite{Giersz11}\\
\hline
($X,Y,Z$)$_\odot$ & ($-8.3,0.0,0.0$)\,kpc & \cite{Gillessen09}\\
$V_\odot$ & $220.0$\,km\,s$^{-1}$ & \cite{Kerr86}\\
($U,V,W$)$_{\rm LSR}$ & (11.1,12.24,7.25)\,km\,s$^{-1}$ &\cite{Schonrich10}\\
\hline
\end{tabular}
\end{center}
\end{table*}
 
\section{Results}\label{results}

Using the input data described in the previous section, we first derive the
orbit of 47 Tuc about the Galactic centre by means of numerical
integration. We then use this orbit to apply the method described by
\cite{Kupper12} for predicting the shape of the tidal tails of the
cluster. Finally we discuss the epicyclic overdensities within those tails.

\subsection{The orbit}\label{orbit}
\begin{figure}
  \begin{centering}
  \includegraphics[width=0.48\textwidth]{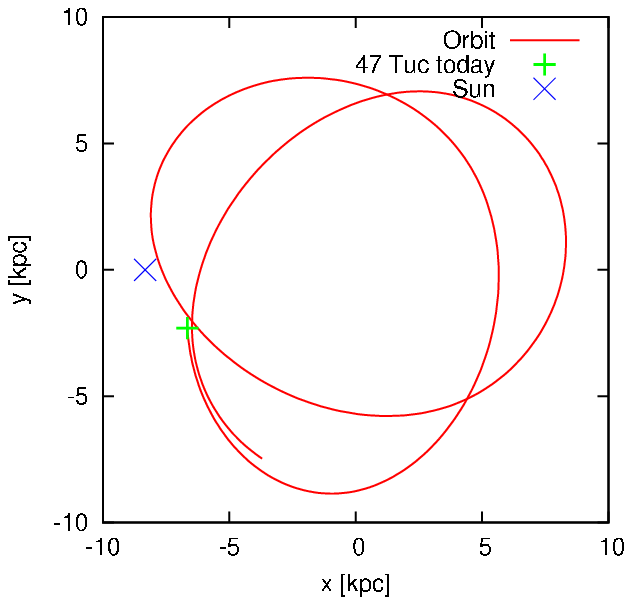}
  \includegraphics[width=0.48\textwidth]{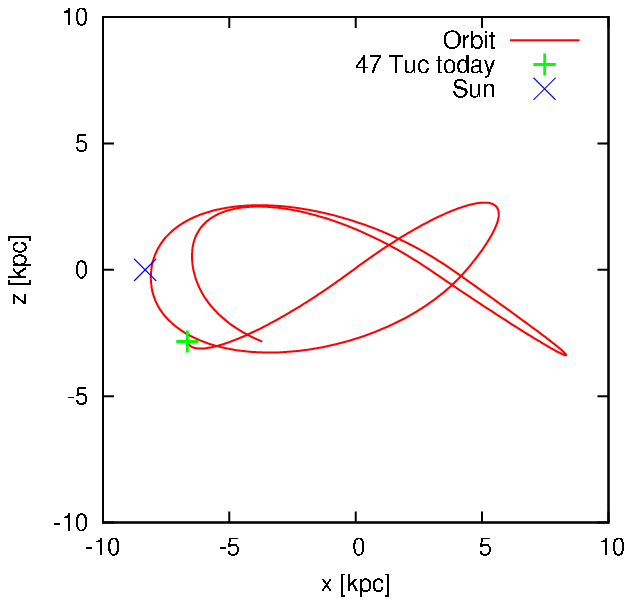}
  \caption{The orbital path of 47 Tuc from 500\,Myr ago to the present time in
    the (x,y) plane (top panel) and in the (x,z) plane (bottom panel), based
    on parameters from the literature (see Table \ref{modelparams}). The
    present day position of 47 Tuc is marked as a plus ($+$) symbol and the
    position of the Sun is marked as a cross ($\times$).}
  \label{orbitxy}
  \end{centering}
\end{figure}
The last 500 Myr of the (x,y) and (x,z) orbital paths of 47 Tuc derived from
the proper motion and radial velocity of the cluster (see Table
\ref{modelparams}) are shown in Fig.~\ref{orbitxy}. For this purpose we
integrated the cluster orbit, assuming the Milky Way potential suggested by
\cite{Allen91}, consisting of a central point mass, a disc and an isothermal
halo. This potential has a circular velocity of 220\,km\,s$^{-1}$ at the
galactocentric distance of the Sun. In total, we integrated the orbit for 1000
Myr back in time, which is the time needed to build up the tidal tails out to
large angular distances from the cluster centre.

The derived orbit yields a perigalactic distance, $R_{peri}=5.90$\,kpc and an
apogalactic distance, $R_{apo}=9.18$\,kpc. Hence, the eccentricity,
$\epsilon$, of the orbit is
\begin{equation}
\epsilon = \frac{R_{apo}-R_{peri}}{R_{apo}+R_{peri}} = 0.22.
\end{equation}
The current galactocentric distance, $R_{GC}$, of 47 Tuc is 7.59\,kpc, and the
cluster is on its way to perigalacticon. The orbital phase, $p_{orb}$, as
defined by \cite{Kupper11a} is, therefore,
\begin{equation}
p_{orb} = \frac{\dot{R}_{GC}}{| \dot{R}_{GC}|}\frac{R_{GC}-R_{peri}}{R_{apo}-R_{peri}} = -0.52,
\end{equation} 
which means that the cluster is half-way to perigalacticon. Its current
orbital velocity is 206\,km\,s$^{-1}$, while it is 160\,km\,s$^{-1}$ at
apogalacticon and 264\,km\,s$^{-1}$ at perigalacticon. Hence, the cluster and
its tails are currently being strongly accelerated.

\subsection{The tidal tails}\label{tails}

For predicting the shape and extent of the tidal tails of 47 Tuc, we apply the
method introduced by \cite{Kupper12}. Starting at 1000 Myr in the past, we
follow the orbit of 47 Tuc and calculate its theoretical tidal radius,
$x_L(t)$, at time $t$ for each timestep using
\begin{equation}\label{eqntdlrad}
x_L (t) = \left( \frac{GM}{\Omega^2-\partial^2\Phi/\partial R^2}\right)^{1/3},
\end{equation}
where $G$ is the gravitational constant, $M$ is the assumed cluster mass,
$\Phi$ is the Galactic potential, $\Omega$ is the angular velocity of the
cluster about the Galactic centre and $R$ is its galactocentric distance
\cite[e.g.][]{Heggie03, Kupper10a}.

Into the combined potential of the Galaxy and the cluster we inject a stream
of test particles, each of which may represent an escaping star. At the time
of its injection each test particle is placed at one of the two Lagrange
points, L1 or L2, which lie at a distance of $\pm x_L$ from the cluster centre
along the galactocentric radius vector of the cluster (the way in which the
initial velocity is chosen is described below). Since the gravitational
attraction of the cluster equals the repulsive forces of the effective
Galactic potential at these points, the test particles can easily escape into
the tidal tails of the cluster. We then integrate each released test particle
to the present time, ignoring the mutual gravitational interaction between the
test particles but taking the gravitational attraction of the cluster into
account.

Since the tidal radius changes rapidly when the cluster goes through
perigalacticon, many test particles can get recaptured on the way to
apogalacticon as the tidal radius quickly grows to its maximal value at
apogalacticon. Therefore, it is no longer practical to release our test
particles from the Lagrange points. Instead we introduce a minimum radius from
which we release test particles, $x_{edge}$, which we denote the edge
radius. That is, the test particles are released from a radial offset, $\Delta
x$, from the cluster given by
\begin{equation}\label{eq:deltax}
\Delta x(t) = \max\left(x_L(t), x_{edge}\right).
\end{equation}
This edge radius is found experimentally by increasing this lower limit from
the perigalactic tidal radius up to the value at which no test particle is
recaptured \cite[see also][for a more detailed
  description]{Kupper12}. However, we will continue to use the phrase
``Lagrange points'' when referring to the points from which we release test
particles.

\begin{figure}
  \begin{centering}
  \includegraphics[width=0.48\textwidth]{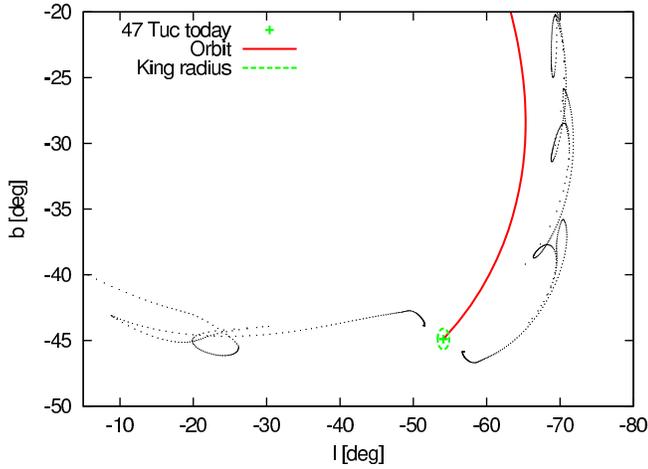}
  \caption{The final timestep (present time) of the reference model is
    shown. The black points are test particles. \new{Due to the manner in
      which they were released from the cluster, these particles can be
      considered streaklines, which allow} for a simple visualisation \new{of
      the positions, at the present time, of stars which have escaped from 47
      Tuc in the recent past.} These have been produced by releasing test
    particles from the tidal radius of the cluster (see text). The solid red
    curve is the orbital path of the cluster. The current position of the
    cluster is marked by a green cross, and its King radius as derived from
    its surface density profile (see Fig.~\ref{densityprofile}) is given by
    the green dashed curve. Escaped stars are slowest within the epicyclic
    loops, which is where overdensities will be visible.}
  \label{cold}
  \end{centering}
\end{figure}

From the results of the numerical integrations we obtain the positions of the
test particles at the present time, giving streaklines such as those plotted
in Fig. \ref{cold}. \new{Streaklines are a concept from fluid dynamics, in
  which test particles are released into a fluid from a given point to
  visualise flow. In engineering streaklines are often produced with smoke or
  dye for tracking flows of air or liquids, respectively \cite[see Section 2.3
    by][for a detailed description of the use of streaklines in the current
    context]{Kupper12}. Specific to the current study, we} let {\bf x}({\it
  t};{\it t}$_0$) be the galactocentric position at time {\it t} of a particle
which was injected at time {\it t}$_0$.  Then a streakline in Fig.~\ref{cold}
is the set of points {\bf x}({\it t};{\it t}$_0$) in which {\it t} is the
present time and {\it t}$_0$ ranges over the duration of the simulation.  Thus
streaklines are not orbits.  By contrast the red curve in Fig.~\ref{cold} is
the orbit of the cluster, the locus of its galactocentric position {\bf
  x}({\it t}), in which {\it t} ranges over the duration of the
simulation. Such plots are common in work on tidal tails, e.g. Fig.~20 by
\cite{Eyre11}.

Our simulations are designed to locate the overdensities in the tidal tails of
47 Tuc caused by the epicyclic motions of the escaping stars (see
Sec.~\ref{overdensities}). \new{These overdensities are the direct result of
  the motions of the escaping stars. Escaped stars have the lowest relative
  motion, with respect to their neighbours, within the epicyclic loops,
  leading to stars `bunching up' at these locations, resulting in
  overdensities.}  We performed several simulations to cover uncertainties in
the cluster mass, proper motion and Heliocentric distance of the
cluster. Therefore, we will first discuss a reference model, then we will vary
single parameters to study the effect of the changes.

Table \ref{modelparams} describes the parameters we have chosen for our
reference model. This reference cluster has a total final mass of $1.0\times
10^6$\,M$_{\odot}$ and loses one star every 0.075\,Myr. Its proper motion is
given by $\mu_\alpha = 5.64$\,mas\,yr$^{-1}$ and $\mu_\delta =
-2.05$\,mas\,yr$^{-1}$ and it has a Heliocentric distance of 4.02\,kpc. We run
our simulations for 1000\,Myr, finishing at the present time. We have chosen
1000\,Myr as this is long enough for the first, second and third order
overdensities in the tidal tails to become fully populated.

In Fig.~\ref{cold} the streaklines of the above setup are shown. To produce
these lines we released the test particles with exactly the angular velocity
of the cluster from exactly the Lagrange points (Eq.~\ref{eq:deltax}). The
edge radius for this cluster mass was found to be 158\,pc, whereas the actual
tidal radius of the cluster varies between 111\,pc at perigalacticon and
171\,pc at apogalacticon. The epicyclic motion of the test particles in the
tails is obvious. The epicyclic loops are strongly influenced and disturbed by
the cluster mass.

\begin{figure}
  \begin{centering}
  \includegraphics[width=0.48\textwidth]{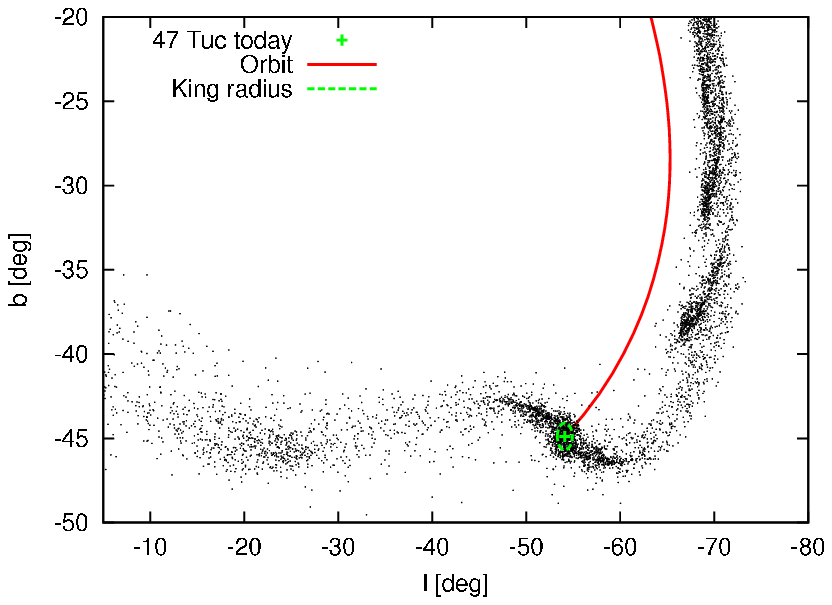}
  \includegraphics[width=0.48\textwidth]{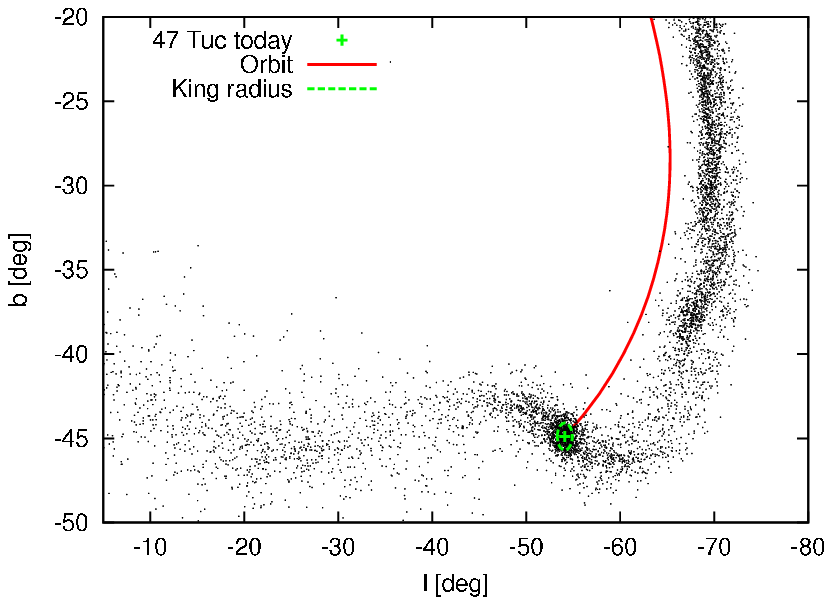}
  \caption{Same as Fig.~\ref{cold}, except with `warm' (top panel) and `hot'
    (bottom panel) escape conditions of the test particles (black points) as
    described in the text. Each test particle has been given a random spatial
    offset and a velocity offset at the moment of escape. Because of these
    offsets the test particles no longer lie along a smooth curve as in
    Fig.~\ref{cold}. The larger the spread in escape conditions, the lower the
    peak density within the epicyclic overdensities, however, the locations of
    the overdensities are only negligibly affected.}
  \label{warm}
  \end{centering}
\end{figure}

In Fig.~\ref{warm} the simulated tails of 47 Tuc are shown for two different
sets of escape conditions. For the `warm' escape conditions (top panel), the
escaping stars retain the angular velocity of the cluster plus they obtain a
random offset in velocity drawn from a Gaussian distribution with a FWHM of
1\,km\,s$^{-1}$. Moreover, a Gaussian offset with a FWHM of 25\% of the tidal
radius has been added about the Lagrange points so that they do not all escape
from a single point on the edge of the cluster. For the `hot' escape
conditions (bottom panel in Fig.~\ref{warm}) we doubled the above fluctuations,
i.e. a Gaussian velocity offset with a FWHM of 2\,km\,s$^{-1}$ and a Gaussian
spatial offset with FWHM of 50\% of the tidal radius. However, from
\cite{Kupper12} it is obvious that the scatter in escape conditions is in
fact not very large. Hence, the `hot' escape conditions can be regarded as the
`worst case scenario'.

We see that adding random fluctuations to the escape conditions of the test
particles increases the width of the tails and scatters their orbits about the
ideal orbits shown in Fig.~\ref{cold}. The larger the scatter in escape
conditions, the smaller the peak density in the epicyclic
overdensities. However, the positions of the overdensities is not altered
significantly.

Furthermore, we have produced several simulations to test how the uncertainty
in the input parameter values affects the output of our model. For proper
motion, we produce models with $\mu_\alpha=4.23$\,mas\,yr$^{-1}$ and
$\mu_\alpha=7.05$\,mas\,yr$^{-1}$ as well as with
$\mu_\delta=-1.54$\,mas\,yr$^{-1}$ and $\mu_\delta=-2.56$\,mas\,yr$^{-1}$. For
our decreased and increased mass models we employ a cluster mass of $0.9\times
10^6$\,M$_{\odot}$ and $1.1\times 10^6$\,M$_{\odot}$, respectively. We also
include Heliocentric distances of 4.7\,kpc and 3.3\,kpc because, during the
writing of this paper, \cite{Woodley12} published a Heliocentric distance of
4.7\,kpc, a discrepancy of about 15\% from the value we assume for our
reference model. For these variations we assumed the same setup as for the
`warm' case mentioned above, except for the parameter which was explicitly
altered. See Table \ref{densitytable} for a summary of the positions and peak
densities of the first and second order epicyclic overdensities in the
trailing tail resulting from each model. These models will be further
discussed in the following section.

\subsection{Epicyclic Overdensities}\label{overdensities}

Despite the density of the epicyclic overdensities changing slightly from the
warm to the hot escape conditions (see Models 1 and 2 in Table
\ref{densitytable}), their positions change very little, therefore, for the
remainder of this paper we will mostly focus on the warm tidal tails, in the
interest of brevity.

\begin{table*}
\begin{center}
\caption{The density and Galactic coordinates of the first and second order
  overdensities, the proper motion, final (present day) cluster mass,
  Heliocentric distance and escape conditions for each of our models. Only the
  trailing tail is represented here for brevity. Information is for the peak
  density in each overdensity only. Note that the approximate average density
  of the stellar background in the regions of the first and second tidal
  overdensities is 3000 stars per square degree, based on 2MASS point
  sources.}\label{densitytable}
\begin{tabular}{@{}cccccccccc@{}}
\hline
\hline
Model & $\Sigma_1$ & $\Sigma_2$ & (l,b)$_1$ & (l,b)$_2$ & $\mu_\alpha$ &
$\mu_\delta$ & $\times10^6$ & $D$ & Esc\\
 & $\star$/deg$^2$ & $\star$/deg$^2$ & deg & deg & mas\,yr$^{-1}$ &
mas\,yr$^{-1}$ & M$_\odot$ & kpc & \\
\hline
1 & 105 & 90 & (-67.2,-38.3) & (-69.2,-31.3) & 5.64 & -2.05 & 1.0 & 4.02 & warm\\
2 & 70 & 55 & (-68.0,-37.9) & (-69.6,-30.3) & 5.64 & -2.05 & 1.0 & 4.02 & hot\\
\hline
3 & 105 & 90 & (-67.4,-37.7) & (-68.2,-30.8) & 7.05 & -2.05 & 1.0 & 4.02 &
warm\\
4 & 115 & 115 & (-68.5,-38.0) & (-69.7,-31.0) & 4.23 & -2.05 & 1.0 & 4.02 & warm\\
\hline
5 & 105 & 95 & (-66.9,-39.7) & (-69.2,-33.6) & 5.64 & -2.56 & 1.0 & 4.02 & warm\\
6 & 115 & 110 & (-67.5,-37.0) & (-69.4,-29.0) & 5.64 & -1.54 & 1.0 & 4.02 & warm\\
\hline
7 & 100 & 90 & (-67.6,-38.1) & (-69.5,-30.9) & 5.64 & -2.05 & 1.1 & 4.02 & warm\\
8 & 110 & 95 & (-67.0,-38.5) & (-69.2,-31.7) & 5.64 & -2.05 & 0.9 & 4.02 & warm\\
\hline
9 & 105 & 100 & (-69.7,-36.0) & (-71.3,-26.6) & 5.64 & -2.05 & 1.0 & 3.30 & warm\\
10 & 130 & 110 & (-65.4,-39.8) & (-67.7,-33.8) & 5.64 & -2.05 & 1.0 & 4.70 & warm\\
\hline
\end{tabular}
\end{center}
\end{table*}

\begin{figure}
  \begin{centering}
  \includegraphics[width=0.48\textwidth]{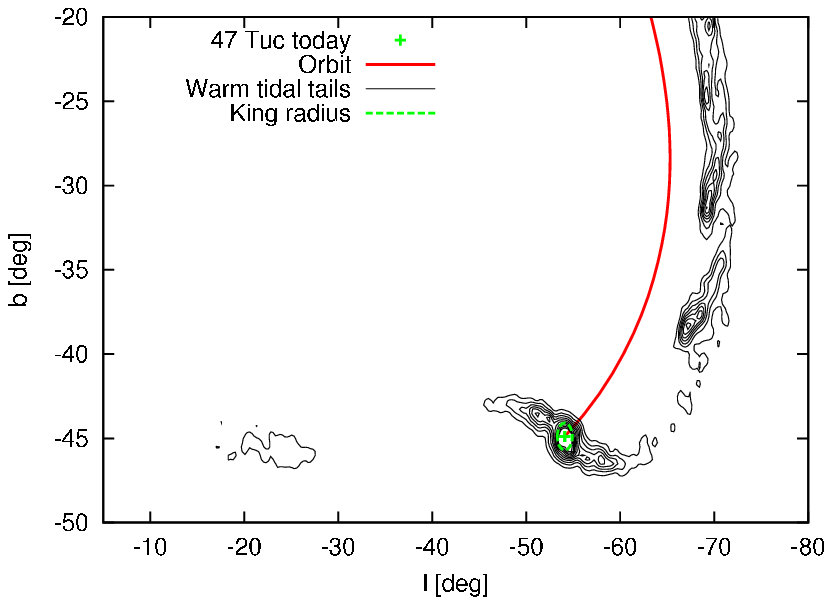}
  \includegraphics[width=0.48\textwidth]{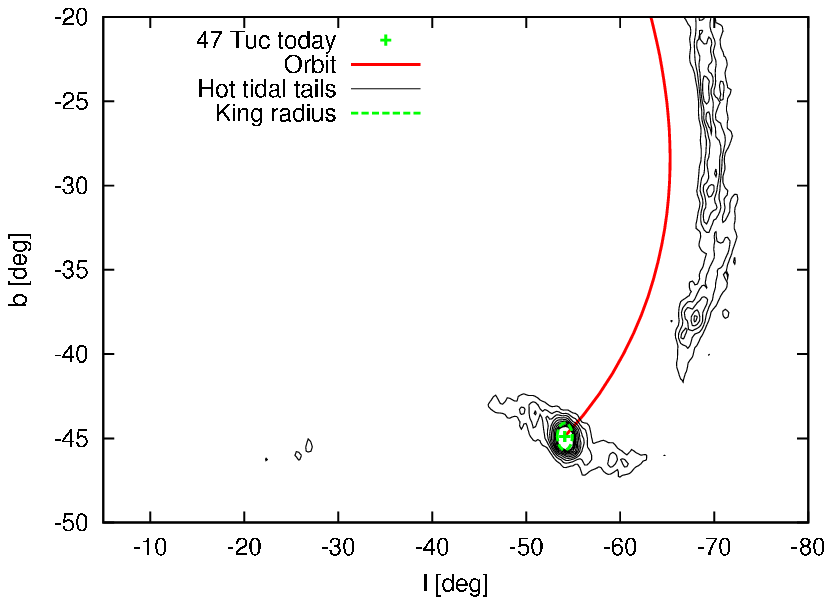}
  \caption{Contour maps of the tidal tails of 47 Tuc as shown in
    Fig.~\ref{warm}. In the top panel the warm escape conditions are shown,
    whereas in the bottom panel the test particles were released with hot
    escape conditions. The contour lines refer to 15, 30, 45, ...,
    stars/deg$^2$. The red solid curve is the orbital path of the cluster. The
    current position of the cluster is marked by a green cross, and its King
    radius as derived from its surface density profile (see
    Fig.~\ref{densityprofile}) is given by the green dashed curve. As can be
    seen in Fig.~\ref{warm}, the peak density in the epicyclic overdensities
    gets much lower when the spread in escape conditions increases.}
  \label{contours1}
  \end{centering}
\end{figure}

\begin{figure}
  \begin{centering}
  \includegraphics[width=0.48\textwidth]{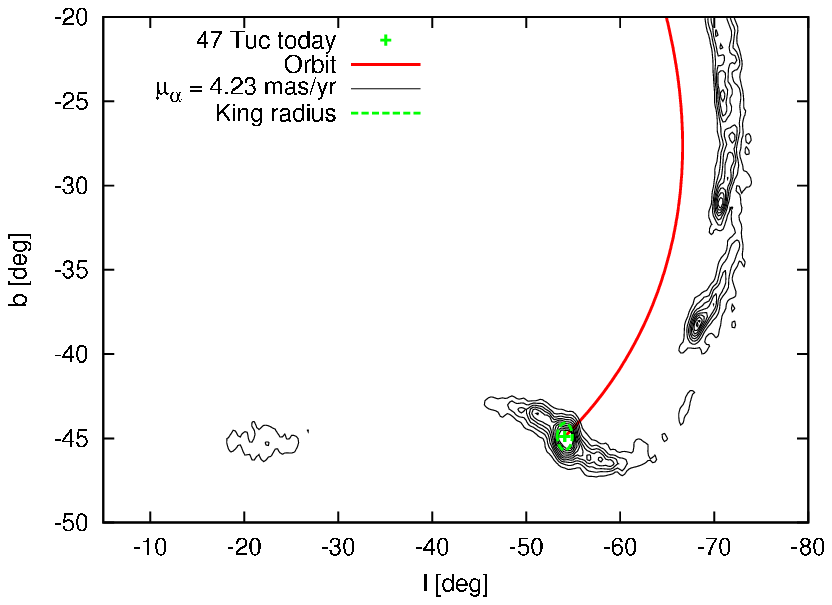}
  \includegraphics[width=0.48\textwidth]{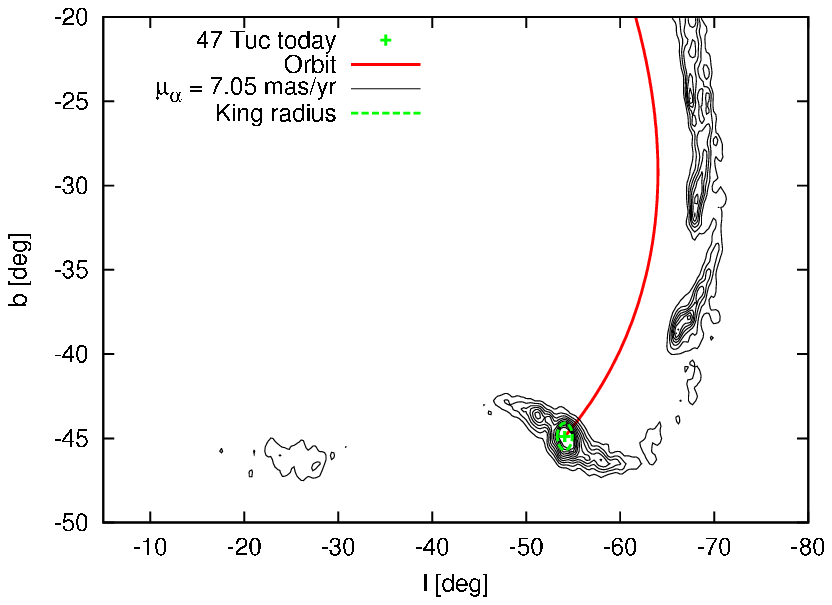}
  \caption{Same as for the top panel of Fig.~\ref{contours1}, except with $\mu_\alpha=4.23$\,mas\,yr$^{-1}$ (top panel) and  $\mu_\alpha=7.05$\,mas\,yr$^{-1}$ (bottom panel).}
  \label{contours2}
  \end{centering}
\end{figure}

\begin{figure}
  \begin{centering}
  \includegraphics[width=0.48\textwidth]{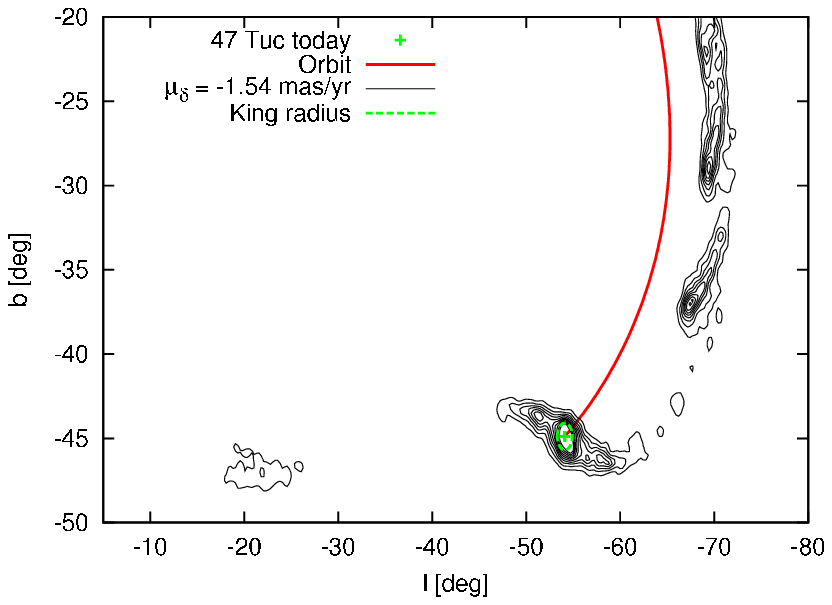}
  \includegraphics[width=0.48\textwidth]{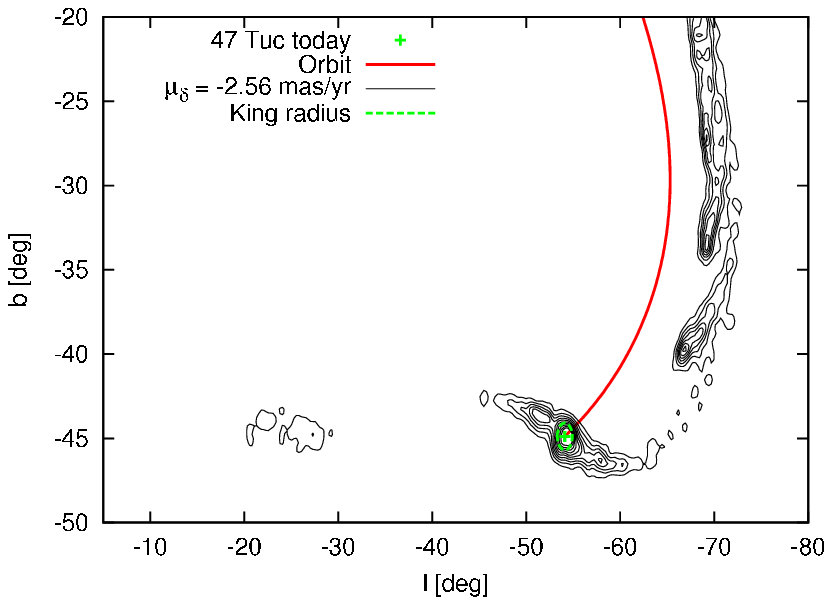}
  \caption{Same as for the top panel of Fig.~\ref{contours1}, except with $\mu_\delta=-1.54$\,mas\,yr$^{-1}$ (top panel) and $\mu_\delta=-2.56$\,mas\,yr$^{-1}$ (bottom panel).}
  \label{contours3}
  \end{centering}
\end{figure}

\begin{figure}
  \begin{centering}
  \includegraphics[width=0.48\textwidth]{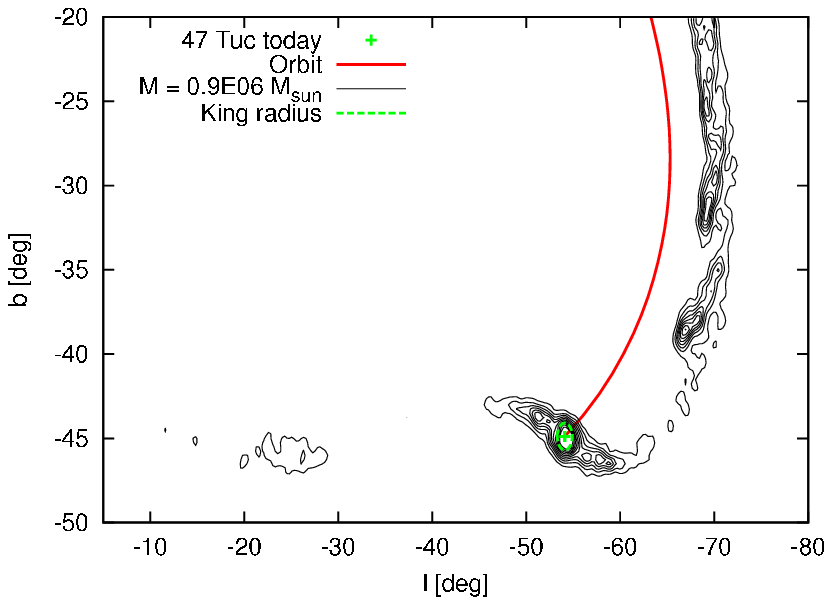}
   \includegraphics[width=0.48\textwidth]{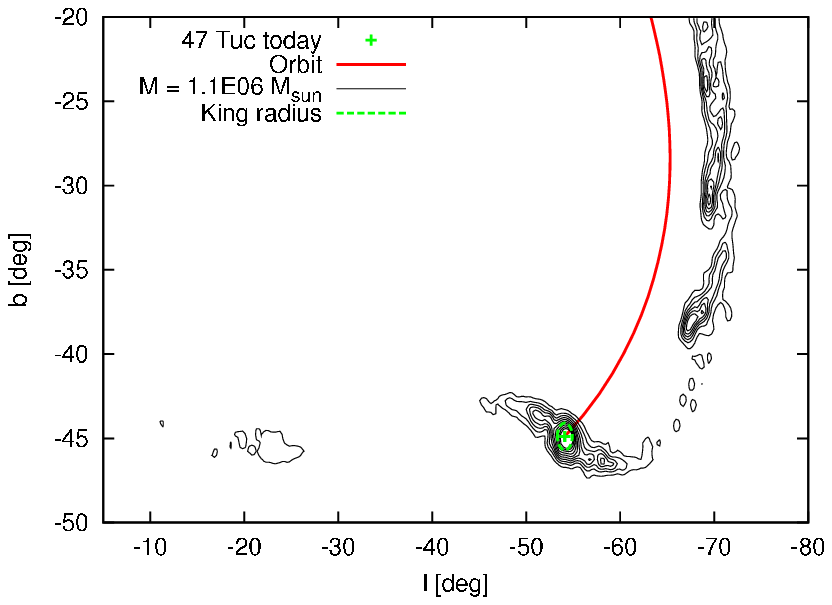}
  \caption{Same as for the top panel of Fig.~\ref{contours1}, except with $M = 0.9\times 10^6$\,M$_{\odot}$ (top panel) and  $M = 1.1\times 10^6$\,M$_{\odot}$ (bottom panel).}
  \label{contours4}
  \end{centering}
\end{figure}

\begin{figure}
  \begin{centering}
  \includegraphics[width=0.48\textwidth]{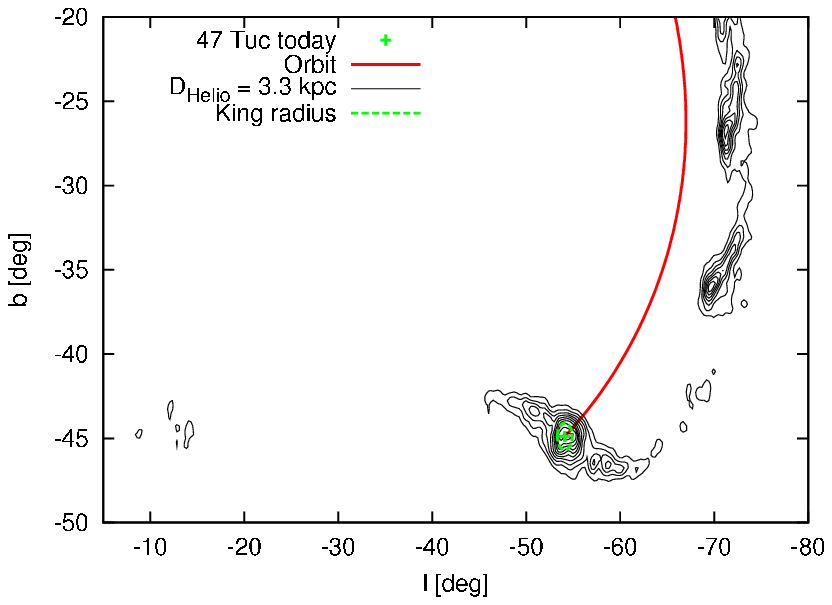}
  \includegraphics[width=0.48\textwidth]{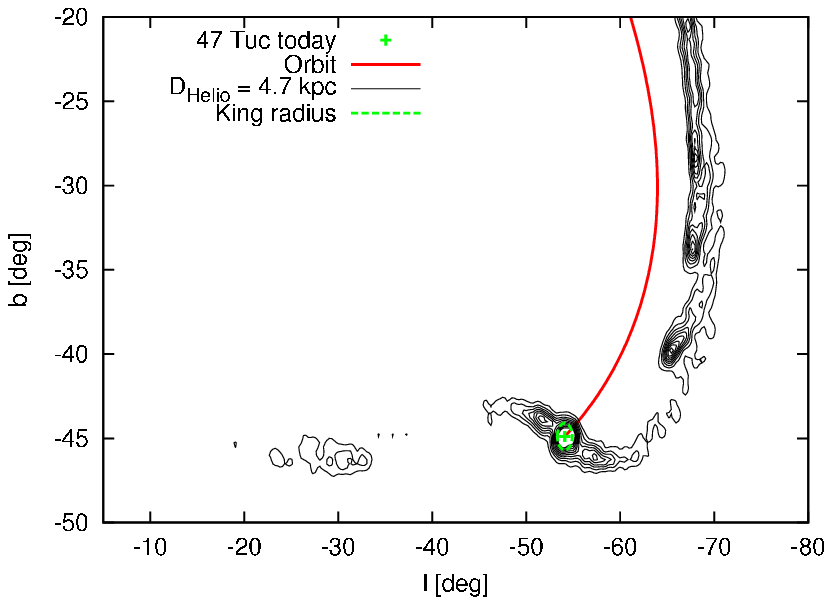}
  \caption{Same as for the top panel of Fig.~\ref{contours1}, except with $D_{helio} = 3.3$\,kpc (top panel) and  $D_{helio} = 4.7$\,kpc (bottom panel).}
  \label{contours5}
  \end{centering}
\end{figure}

Density contours of the warm tidal tails can be seen in Figures
\ref{contours1} to \ref{contours5}. The epicyclic overdensities exhibit
surface densities $\sim3-4\%$ above the average Galactic background in the
region near the tidal tails, depending on the model (see Table
\ref{densitytable}). Tidal features exhibiting very low surface brightness
variations are recoverable using matched filtering techniques. For example,
\cite{Grillmair06} resolved tidal tails with extremely low surface brightness
over a region of sky with a mean stellar background of $\sim4500$\,deg$^{-2}$,
about 50\% higher than the background near the tidal overdensities of 47
Tuc. Furthermore, after applying an ``optimal contrast filter'' (effectively a
matched filter), \cite{Odenkirchen03} detected features with a peak $\sim20\%$
above the background counts, and much smaller deviations from the background
between the density peaks, for the tidal tails of the globular cluster Pal
5. Indeed, the authors claim that the density peaks are about twice the mean
density of the tidal tails overall (i.e. the mean observed density of the
tails is 10\% above the background counts). Assuming their claimed 4.3 times
increase in contrast using the matched filter, the density peaks for the
epicyclic overdensities in the tidal tails of 47 Tuc will be $\gtrsim13\%$
above the background, well above the detection threshold.

Furthermore, it can be seen that altering various parameters also affects the
positions of the first and second order overdensities differently. This is due
to the different influence of the cluster mass and the orbital phase on the
appearance of the tidal tails and the overdensities
\cite[see][]{Kupper10a,Kupper12}. Hence, not only the location of the
overdensities with respect to the cluster is of interest when constraining the
dynamical state of the cluster, but also the relative distance between the
overdensities themselves.

\begin{figure}
  \begin{centering}
  \includegraphics[width=0.48\textwidth]{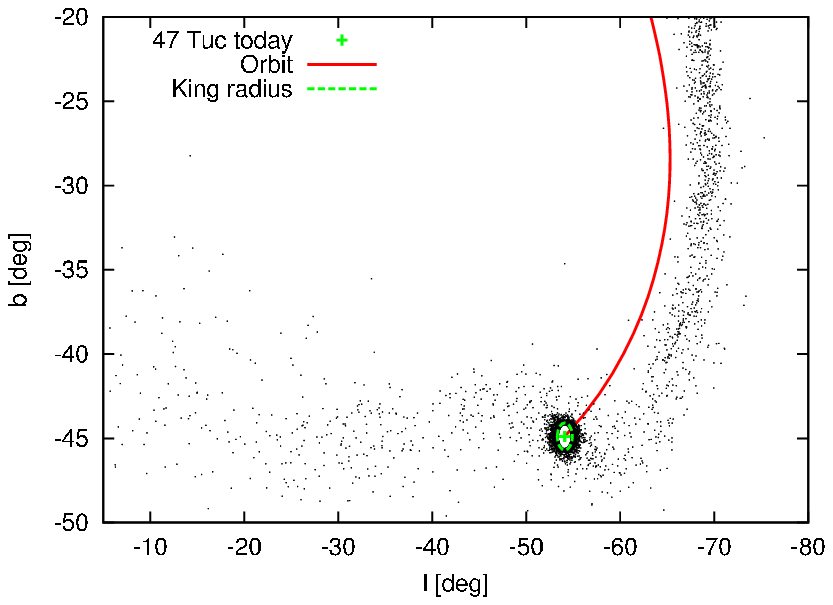}
  \includegraphics[width=0.48\textwidth]{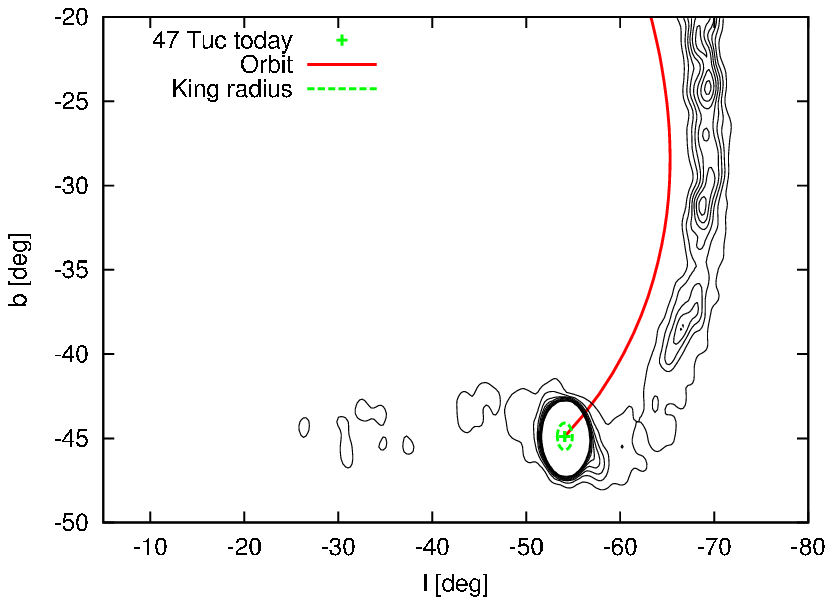}
  \caption{Distribution of tail stars from a simple $N$-body simulation. The
    shape of the tails and the positions of the overdensities agree very well
    with the predictions from our streakline models. A more realistic $N$-body
    model with a larger number of particles (here we use \new{110\,000}) will
    show less scatter in escape conditions and will result in dynamically
    colder tidal tails (see Sec. \ref{nbodysec}). Particles within the King
    radius are not shown in the upper panel for clarity. The contour lines in
    the lower panel refer to 2.5, 5, 7.5, ..., particles per square
    degree. Due to the smaller number of particles in the tidal tails we used
    a Gaussian smoothing kernel with FWHM~$=1^\circ$, twice as large as for
    our other contour plots (Figs.~\ref{contours1}-\ref{contours5}).}
  \label{nbody}
  \end{centering}
\end{figure}

\subsection{Comparison with $N$-body simulations }\label{nbodysec}

Figures 7-10 by \cite{Kupper12} show that our formalism for releasing test
particles into the tidal tails yields very good agreements with results from
direct $N$-body simulations. However, in comparison to the models presented in
that study, 47 Tuc moves on a slightly more complex orbit which also includes
periodic disc crossings. Depending on where a cluster crosses the Galactic
disc, disc shocks can significantly heat up the cluster and increase its
overall mass-loss rate \cite[e.g.][]{Vesperini97}. Furthermore,
Eq.~\ref{eqntdlrad}, which we use for calculating the distance from which we
release the test particles, has to be taken as a rough approximation during
these events. Therefore, we have to check if our assumptions are still valid
under these conditions.

Unfortunately, a direct $N$-body simulation of 47 Tuc is currently not
feasible due to the large number of member stars (about $2\times10^6$).
Instead, we set up a \new{moderately high resolution} toy model of 47 Tuc with
\new{$1.1\times10^5$} particles using the publicly available code
\textsc{McLuster}\footnote{\texttt{www.astro.uni-bonn.de/\~{}akuepper/mcluster/mcluster.html\\www.astro.uni-bonn.de/\~{}webaiub/german/downloads.php}}
\cite[][]{Kupper11b}. We used a spectrum of particle masses, $m$, ranging from
$4-40\,\mbox{M}_\odot$ and a mass distribution proportional to $m^{-2.35}$,
adding up to a total mass of \new{$1.0\times10^6\,\mbox{M}_\odot$}. The
particles were distributed in a \cite{Plummer11} sphere with a half-mass
radius of 10\,pc, largely consistent with the surface density profile of 47
Tuc. The cluster was integrated along the orbit of 47 Tuc for 1\,Gyr using
\textsc{Nbody6} \cite[][]{Aarseth03} on a GPU computer at AIfA Bonn. Over the
course of the simulation the cluster model lost about 3\% of its mass into the
tidal tails such that its final mass resembles the present-day mass of 47 Tuc,
within observational uncertainties.

In Fig.~\ref{nbody} the cluster and its tails are shown after the
integration. Given the simplicity of our $N$-body model, the distribution of
particles in the tails resembles our predictions from the streakline models
very well ({\it cf.}~Figs.~\ref{warm} \& \ref{contours1}). The overall shape
and the positions of the overdensities matches the predictions from the `hot'
escape conditions best. However, a more realistic simulation with $\sim20$
times the number of particles would have a longer relaxation time and,
therefore, its mass-loss rate would differ from that of our model. Hence, the
particle density in this simulation cannot be taken as direct proxy for the
expected stellar density. Moreover, a simulation with $2\times10^6$ particles
would exhibit much less scatter in escape conditions and, therefore,
dynamically colder tidal tails \cite[see][]{Kupper08}. Thus, we expect that
the tidal tails of 47 Tuc rather resemble the `warm' escape conditions.

\new{Our simplified model (Section \ref{tails}) assumes that the injection
  rate is constant, however, in the $N$-body model described in the current
  Section, the Galactic potential includes a disc component. This may, in
  principle, modulate the escape rate via disc shocking.  However,} this
simulation demonstrates that disc shocks do not significantly affect the shape
of the tidal tails or the positions of the \new{epicyclic} overdensities
within them.  \new{Furthermore, while it is true that the additional energy
  input from tidal shocks, which is distributed within the cluster, results in
  a higher average mass loss rate, it was estimated by \cite{Aguilar88} that
  such shocks only contribute $\sim1\%$ to the total destruction
  rate. Moreover, this increase in mass loss is distributed nearly
  homogeneously over the whole orbit because the escape of the stars is
  delayed until well after the disc shock, often for several orbits of the
  star within the cluster, even for clusters on non-circular orbits
  \cite[][]{Kupper10a}. This delay is the result of the stars preferentially
  leaving the system through the Lagrange points, where the potential barrier
  is at its minimum. Stars excited to escape energies by shocks leave the
  cluster on about the same time-scale as due to other evaporative processes,
  and may even stay bound to the cluster for as long as a Hubble time
  \cite[][]{Fukushige00}, at least for clusters on circular orbits. The
  energetically unbound stars enhancing the outer velocity dispersion of 47
  Tuc \cite[][]{Lane10a} also support the concept of delayed escape,
  especially if these unbound stars were excited during the last disc passage
  which, from our simulations, occurred about 70\,Myr ago. Therefore, because
  energetically unbound stars cannot all escape immediately following shocks,
  and can stay bound to the cluster for long periods, we do not expect to see
  additional substructure in the tidal tails caused by disc crossings.}


While we used the analytic potential suggested by \cite{Allen91} for the
streakline models as well as for the $N$-body model, \cite{Dehnen98} argue
that the Galactic disc structure is actually more complex than the
single-component exponential disc used by Allen \& Santillan. However, the
strongest disc shock 47 Tuc experiences is at 6\,kpc. At this distance the two
potentials do not differ greatly \cite[e.g.][]{Kaempf05,Jilkova12}. Therefore,
we do not expect our predictions to be significantly affected by the stronger
disc shocking suggested by \cite{Dehnen98}.

\section{Conclusions}\label{conclusions}

We have employed the most accurate available data on the Milky Way globular
cluster 47 Tuc to compute its orbit about the Galaxy. Moreover, we used the
streakline method \cite[][]{Kupper12} to predict the shape of the tidal tails
of 47 Tuc and the locations of the epicyclic overdensities within those
tails. By varying the input parameters of our models we show how the most
important input values influence the locations and peak surface densities of
the epicyclic overdensities.

We find that the epicyclic overdensities exhibit peak surface densities
$\sim3-4\%$ above the Galactic background. These should be observationally
detectable, assuming a matched filter, or similar analysis technique, is used.

Clearly variations in Heliocentric distance and proper motion alter the
positions of the tidal overdensities more than variations in any other
parameter in our model. Our variations in proper motion are much larger than
the observational error stated by \cite{Anderson04} and we find only a small
change in the position of the overdensities. However, determining the distance
to 47 Tuc is an ongoing problem, with many different estimates published over
many years \cite[see][for several examples]{Gratton03}. Therefore, finding
these overdensities observationally will not only provide further constraints
on the proper motion of the cluster, but will also constrain the other
dynamical characteristics, including, importantly, the Heliocentric distance.

In addition, uncovering these overdensities observationally may reinforce our
understanding of Newtonian gravity. While it is not within the scope of the
current paper to make predictions regarding the tidal tails and their
overdensities in modified gravity environments, it would be extremely
interesting to see the morphology of the tidal tails as predicted by modified
gravitational theories, since we would expect them to differ significantly
from our predictions.

Finally, our study has shown how, in principle, the shape and extent of the
tidal tails of any star cluster can be quickly predicted without requiring
extensive $N$-body simulations. This will be especially helpful for large
parameter studies or for massive star clusters which cannot be addressed by
means of direct $N$-body computations due to hardware limitations.


\section*{Acknowledgments} 

R.R.L. gratefully acknowledges support from the Chilean Center for
Astrophysics, FONDAP No. 15010003. A.H.W.K. kindly acknowledges support by the
ESO Director General Discretionary Fund and the German Research Foundation
(DFG) project KR 1635/28-1.

Many thanks go to L\'aszl\'o Kiss for the use of the data that went into this
publication, including velocities extracted from the raw spectra.

\new{We would like to thank the anonymous referee for many useful comments.}

\bsp 
 
\label{lastpage} 
 
\end{document}


\date{Accepted..... Received.....; in original form.....} 
 
\pagerange{\pageref{firstpage}--\pageref{lastpage}} \pubyear{2012} 
 
\maketitle 
 
\label{firstpage} 
 
 
\begin{keywords} 
Galaxy: globular clusters: general -- Galaxy: globular clusters: individual:
47 Tucanae -- Galaxy: kinematics and dynamics -- methods: numerical
\end{keywords} 

The paper ``The Tidal Tails of 47 Tucanae'' was published in Mon. Not. R.
Astron. Soc. 423, 2845 (2012). An error was discovered, which affected the
results reported in that paper. This error is corrected here.

We used, for our proper motion of 47 Tucanae (47 Tuc), the values published by
\cite{Anderson04}, however, we incorrectly adjusted the published values of
$\mu_\alpha$ by $\cos{\delta}$. These values were already published with the
$\cos{\delta}$ included so this was unnecessary. Our new reference model for
this erratum uses proper motions:
\begin{eqnarray}
\mu_\alpha\cos{\delta}&=&5.64\,\mbox{mas}\,\mbox{yr}^{-1}\\
\mu_\delta&=&-2.05\,\mbox{mas}\,\mbox{yr}^{-1}
\end{eqnarray}
\noindent All other input parameters were correct and are left unchanged. All
references for, and reasons behind, our choices for the input parameters can
be found in the original paper. The newly computed models are presented in the
following.

\begin{table}
\begin{center}
\caption{Revised orbital parameters of the reference model}\label{newparams}
\begin{tabular}{@{}cc@{}}
\hline
\hline
Parameter & Value\\
\hline
Edge radius & $142$\,pc\\
Minimum orbital velocity & 179\,km\,s$^{-1}$\\
Maximum orbital velocity & 254\,km\,s$^{-1}$\\
Current orbital velocity & 184\,km\,s$^{-1}$\\
Apocentre & $7.65$\,kpc\\
Pericentre & $6.10$\,kpc\\
Current Galactocentric distance & $7.60$\,kpc\\
Orbital eccentricity & 0.11\\
Orbital phase & $+0.97$\\
\hline
\end{tabular}
\end{center}
\end{table}

\section{The orbit}\label{orbit}

\begin{figure}
  \begin{centering}
  \includegraphics[width=0.48\textwidth]{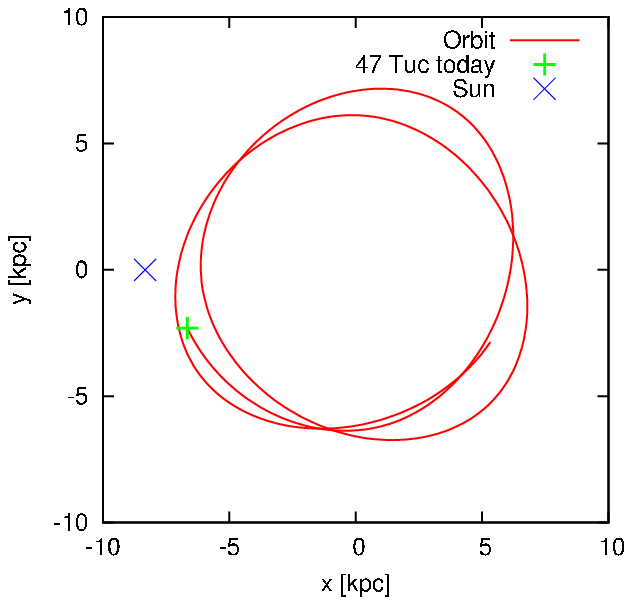}
  \includegraphics[width=0.48\textwidth]{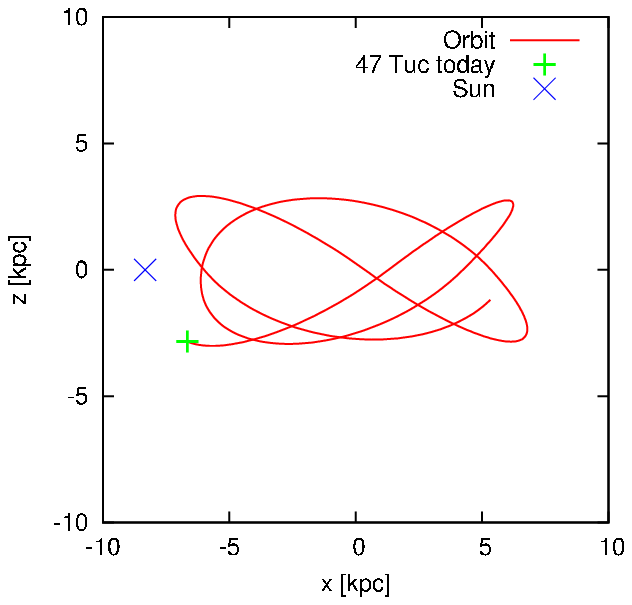}
  \caption{The orbital path of 47 Tuc from 500\,Myr ago to the present time in
    the (x,y) plane (top panel) and in the (x,z) plane (bottom panel), based
    on parameters from the corrected proper motions (see text). The present
    day position of 47 Tuc is marked as a plus ($+$) symbol and the position
    of the Sun is marked as a cross ($\times$).  Corresponds to Figure 2
    of the original paper.}
  \label{orbitxy}
  \end{centering}
\end{figure}

The last 500 Myr of the (x,y) and (x,z) orbital paths of 47 Tuc derived from
the new proper motion and radial velocity of the cluster are shown in
Figure~\ref{orbitxy}. Due to the significant change in proper motion, various
orbital parameters have been altered. The new quantities, defined as per the
original paper, are listed in Table \ref{newparams}. We particularly note that
(for our choice of Galaxy model) the orbit of 47 Tuc is less eccentric than
originally thought ($\epsilon = 0.11$) and that 47 Tuc is very close to
reaching the apocentre of its orbit ($p_{orb} = +0.97$).

\section{The tidal tails}\label{tails}

\begin{figure}
  \begin{centering}
  \includegraphics[width=0.48\textwidth]{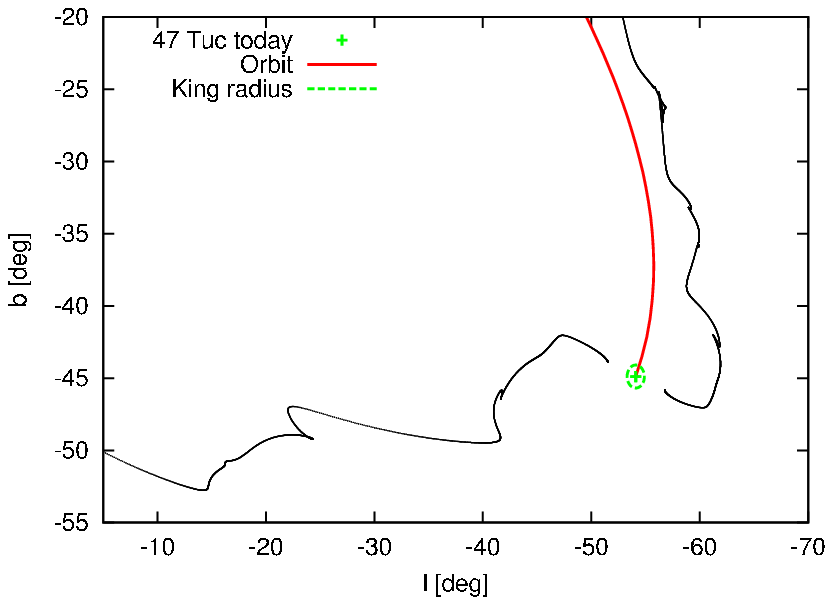}
  \caption{The final timestep for the reference model. Corresponds to Figure 3
    from the original paper.}
  \label{cold}
  \end{centering}
\end{figure}

\begin{figure}
  \begin{centering}
  \includegraphics[width=0.48\textwidth]{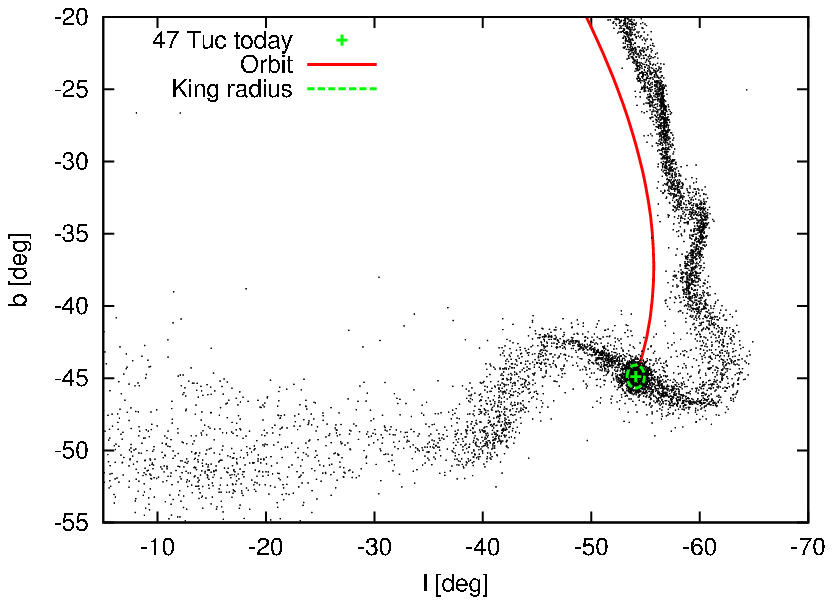}
  \includegraphics[width=0.48\textwidth]{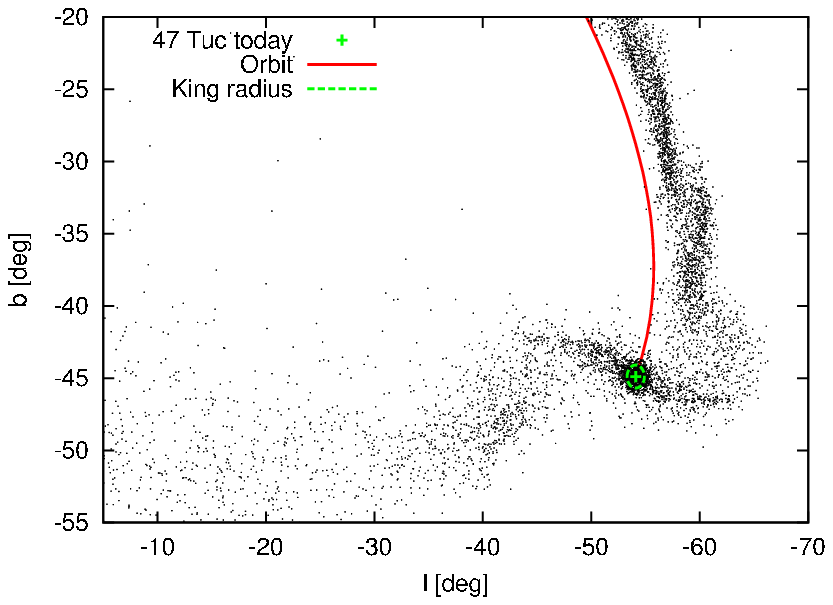}
  \caption{Same as Figure \ref{cold} except with warm (top) and hot (bottom)
    escape conditions. Corresponds to Figure 4 of the original paper.}
  \label{warm}
  \end{centering}
\end{figure}

\begin{figure}
  \begin{centering}
  \includegraphics[width=0.48\textwidth]{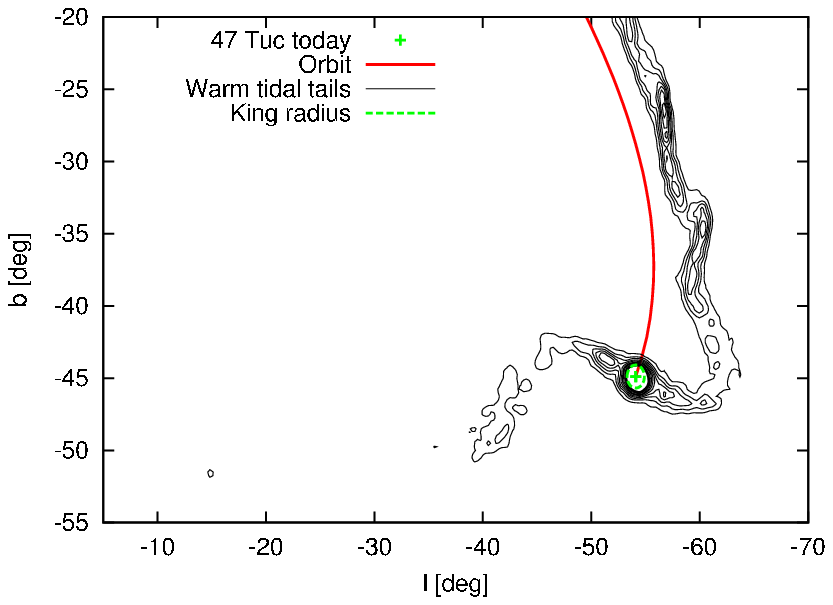}
  \includegraphics[width=0.48\textwidth]{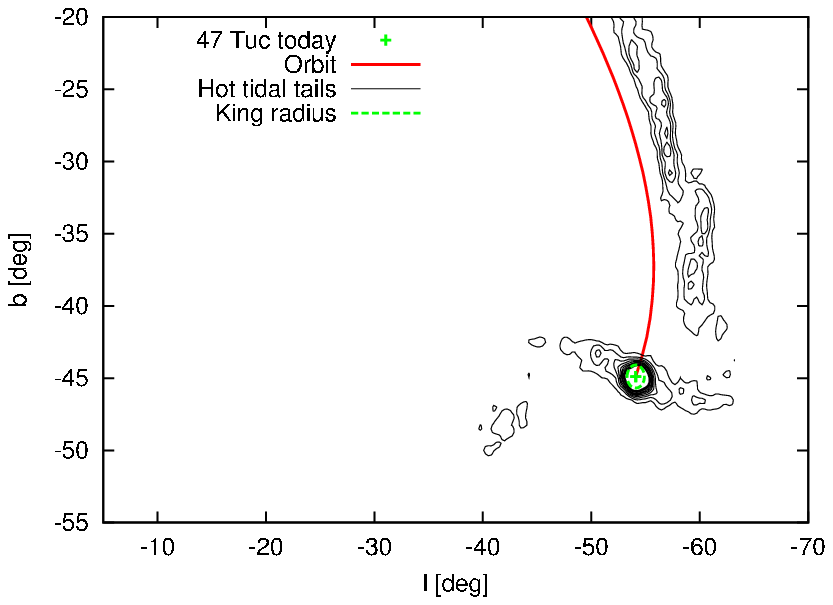}
  \caption{Contour representation of Figure \ref{warm} with ``warm'' escape
    conditions in the top panel and ``hot'' escape conditions in the bottom
    panel. Corresponds to Figure 5 of the original paper.}
  \label{contours1}
  \end{centering}
\end{figure}

\begin{figure}
  \begin{centering}
  \includegraphics[width=0.48\textwidth]{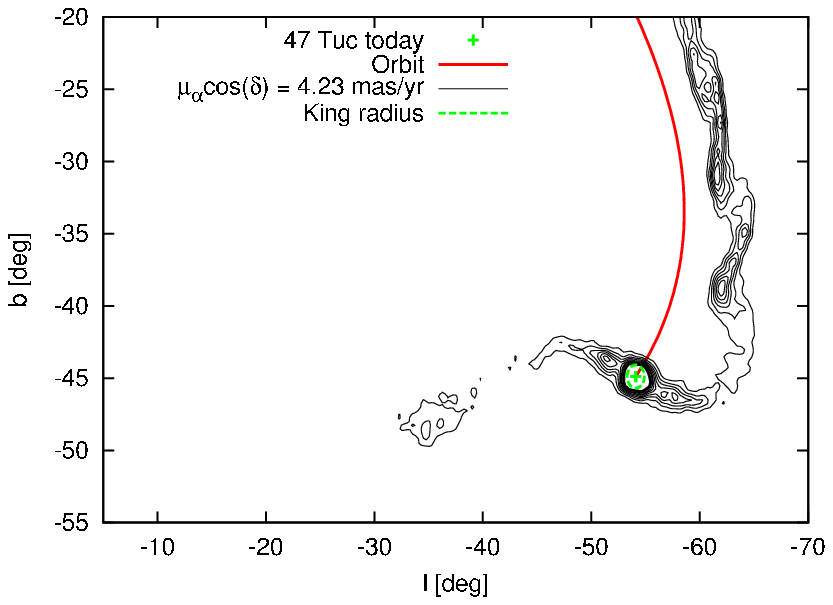}
  \includegraphics[width=0.48\textwidth]{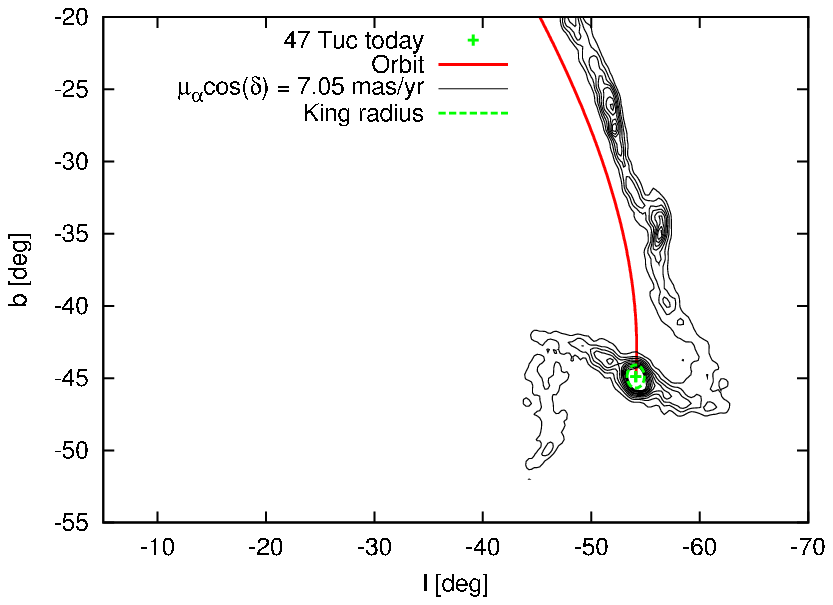}
  \caption{Same as for the top panel of Figure~\ref{contours1}, except with
    $\mu_\alpha\cos{\delta}=4.23$\,mas\,yr$^{-1}$ (top panel) and
    $\mu_\alpha\cos{\delta}=7.05$\,mas\,yr$^{-1}$ (bottom panel). Corresponds
    to Figure 6 of the original paper.}
  \label{contours2}
  \end{centering}
\end{figure}

\begin{figure}
  \begin{centering}
  \includegraphics[width=0.48\textwidth]{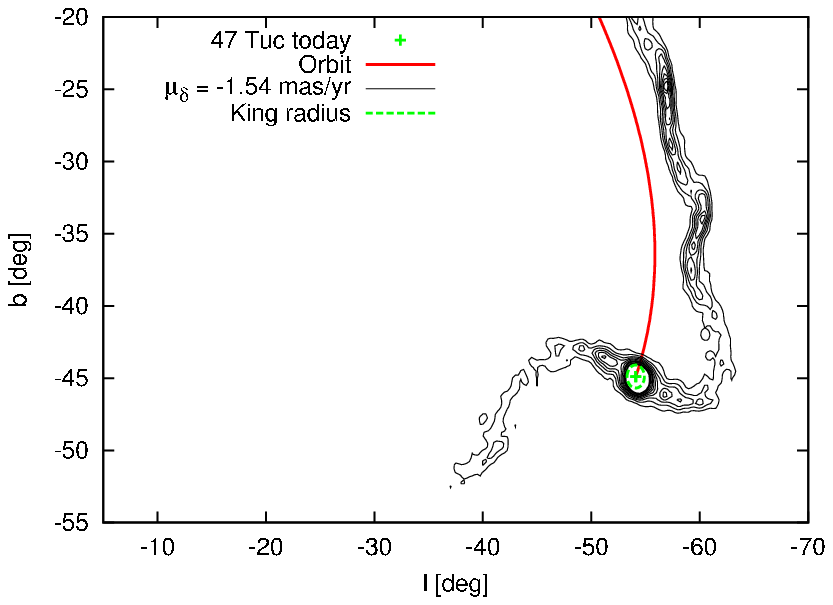}
  \includegraphics[width=0.48\textwidth]{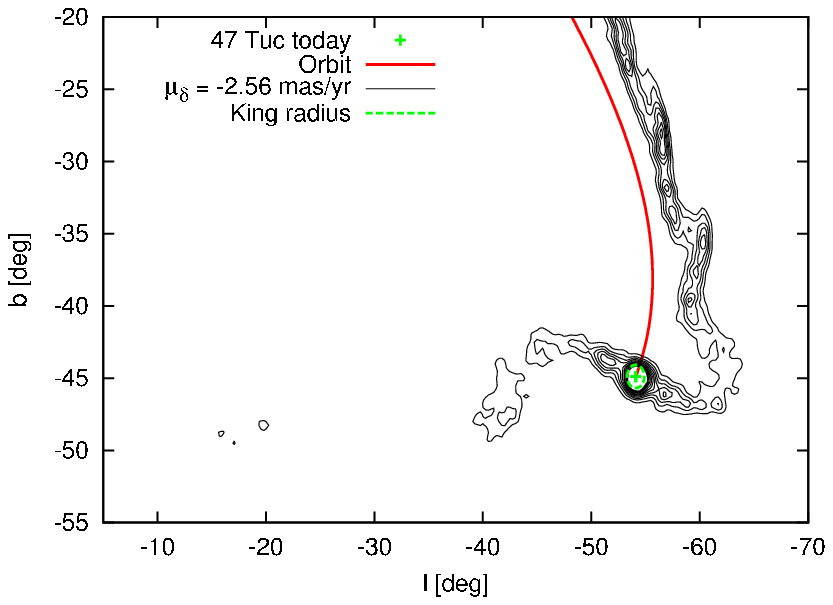}
  \caption{Same as for the top panel of Figure~\ref{contours1}, except with
    $\mu_\delta=-1.54$\,mas\,yr$^{-1}$ (top panel) and
    $\mu_\delta=-2.56$\,mas\,yr$^{-1}$ (bottom panel). Corresponds to Figure 7
    of the original paper.}
  \label{contours3}
  \end{centering}
\end{figure}

\begin{figure}
  \begin{centering}
  \includegraphics[width=0.48\textwidth]{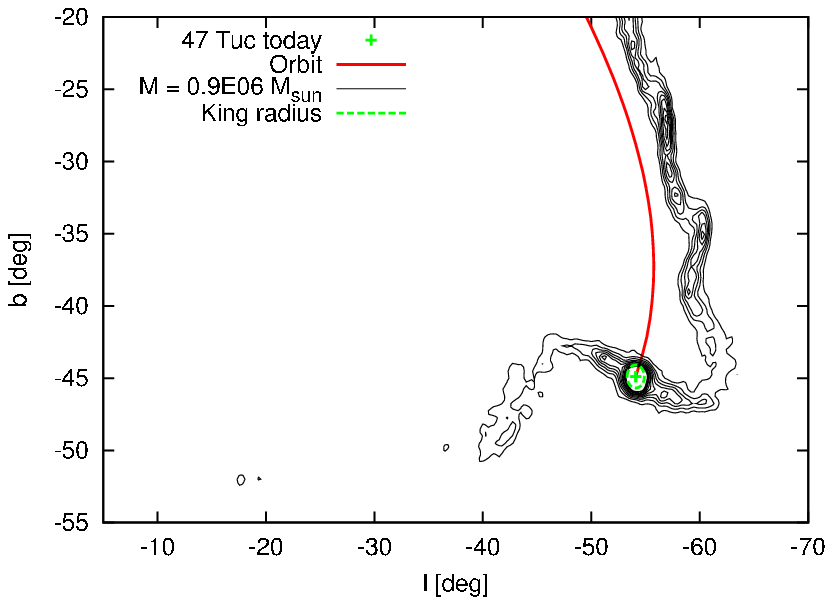}
   \includegraphics[width=0.48\textwidth]{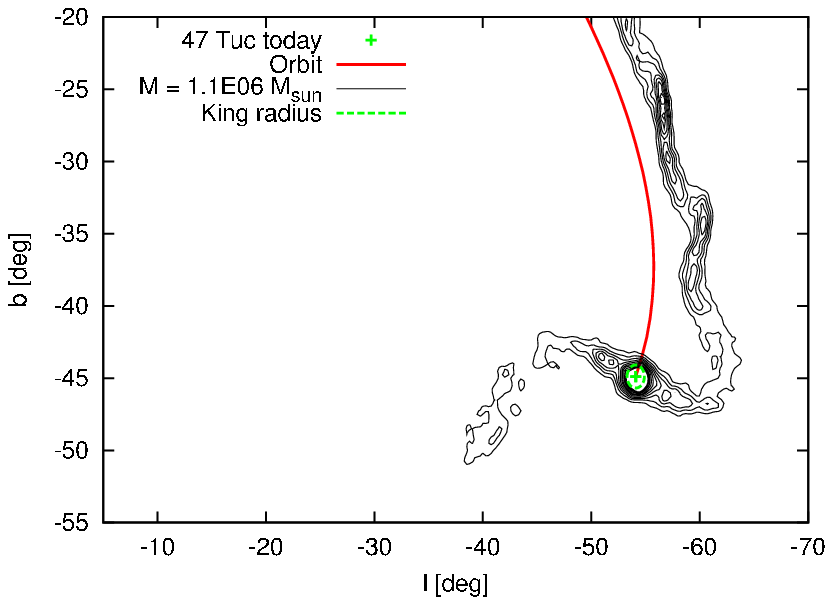}
  \caption{Same as for the top panel of Figure~\ref{contours1}, except with $M
    = 0.9\times 10^6$\,M$_{\odot}$ (top panel) and $M = 1.1\times
    10^6$\,M$_{\odot}$ (bottom panel). Corresponds to Figure 8 of the original
    paper.}
  \label{contours4}
  \end{centering}
\end{figure}

\begin{figure}
  \begin{centering}
  \includegraphics[width=0.48\textwidth]{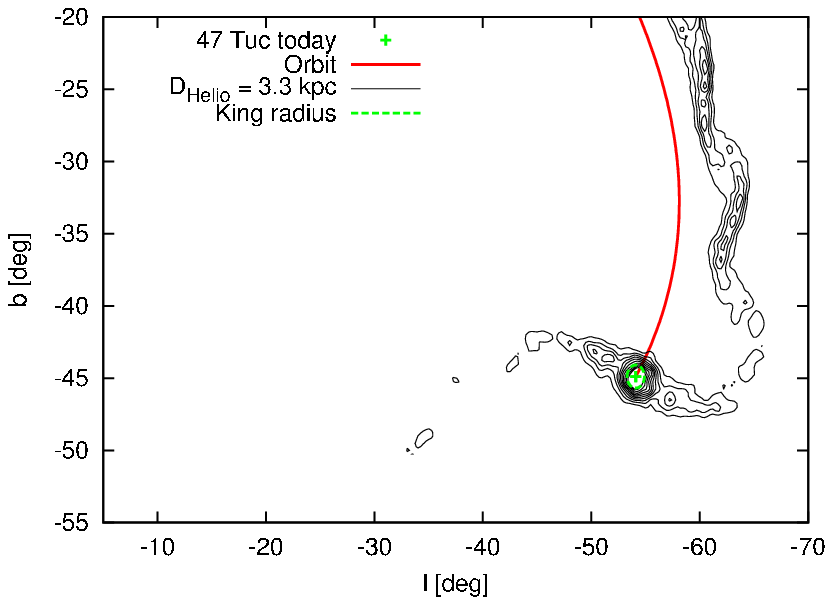}
  \includegraphics[width=0.48\textwidth]{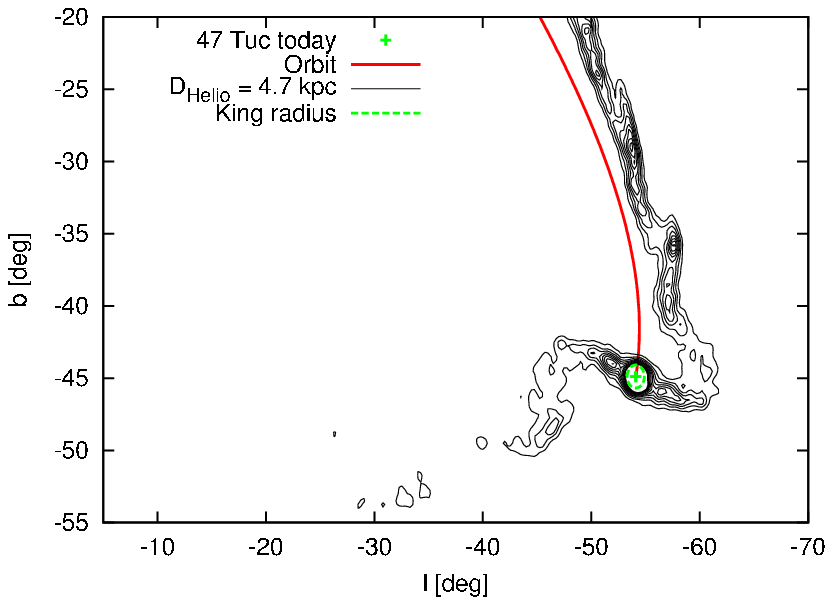}
  \caption{Same as for the top panel of Figure~\ref{contours1}, except with
    $D_{helio} = 3.3$\,kpc (top panel) and $D_{helio} = 4.7$\,kpc (bottom
    panel). Corresponds to Figure 9 of the original paper.}
  \label{contours5}
  \end{centering}
\end{figure}

Figures~\ref{cold}-\ref{contours5} in this erratum are based on our new
models, and correspond to Figures~3-9 from the original
paper. Table~\ref{densitytable} summarises our corrected
results. Qualitatively, the results are effectively equivalent, but the shape
of the tidal tails and, therefore, the positions of their overdensities have
changed.

!

!

!

!

!

!

!

!

!

!

\begin{table*}
\begin{center}
\caption{The density and Galactic coordinates of the first and second order
  overdensities, the proper motion, final (present day) cluster mass,
  Heliocentric distance and escape conditions for each of our corrected
  models. Only the trailing tail is represented here for brevity. Information
  is for the peak density in each overdensity only. Note that the approximate
  average density of the stellar background in the regions of the first and
  second tidal overdensities is 3000 stars per square degree, based on 2MASS
  point sources.}\label{densitytable}
\begin{tabular}{@{}cccccccccc@{}}
\hline
\hline
Model & $\Sigma_1$ & $\Sigma_2$ & (l,b)$_1$ & (l,b)$_2$ & $\mu_\alpha\cos{\delta}$ &
$\mu_\delta$ & $\times10^6$ & $D$ & Esc\\
 & $\star$/deg$^2$ & $\star$/deg$^2$ & deg & deg & mas\,yr$^{-1}$ &
mas\,yr$^{-1}$ & M$_\odot$ & kpc & \\
\hline
1 & 85 & 115 & (-60.4,-34.5) & (-56.7,-26.5) & 5.64 & -2.05 & 1.0 & 4.02 & warm\\
2 & 55 & 70 & (-60.5,-34.5) & (-57.0,-29.2) & 5.64 & -2.05 & 1.0 & 4.02 & hot\\
\hline
3 & 120 & 110 & (-56.5,-34.9) & (-52.2,-27.7) & 7.05 & -2.05 & 1.0 & 4.02 & warm\\
4 & 100 & 100 & (-62.0,-38.7) & (-61.6,-31.0) & 4.23 & -2.05 & 1.0 & 4.02 & warm\\
\hline
5 & 85 & 110 & (-56.5,-28.0) & (-69.2,-33.6) & 5.64 & -2.56 & 1.0 & 4.02 & warm\\
6 & 95 & 130 & (-60.3,-34.0) & (-56.9,-24.7) & 5.64 & -1.54 & 1.0 & 4.02 & warm\\
\hline
7 & 85 & 120 & (-60.4,-34.3) & (-56.5,-25.5) & 5.64 & -2.05 & 1.1 & 4.02 & warm\\
8 & 95 & 130 & (-60.2,-35.0) & (-57.0,-27.0) & 5.64 & -2.05 & 0.9 & 4.02 & warm\\
\hline
9 & 65 & 85 & (-63.7,-32.6) & (-60.4,-23.5) & 5.64 & -2.05 & 1.0 & 3.30 & warm\\
10 & 120 & 125 & (-57.7,-35.7) & (-53.7,-29.0) & 5.64 & -2.05 & 1.0 & 4.70 & warm\\
\hline
\end{tabular}
\end{center}
\end{table*}

\section*{Acknowledgments} 

The authors gratefully thank Radek Poleski for bringing the error to our
attention.

\bsp 
 
\label{lastpage}